\documentclass[reprint,amsmath,amssymb,aps,prd]{revtex4-1}
\usepackage{bm}
\usepackage{dcolumn}
\usepackage{wrapfig}
\usepackage{lineno}
\usepackage{graphicx}
\usepackage{subfigure}
\usepackage{epstopdf}
\usepackage{float}
\usepackage{threeparttable}
\usepackage{overpic}
\usepackage{subfigure}
\usepackage{float}
\usepackage{cmap} 
\usepackage{microtype} 
\usepackage{mathtext}
\usepackage{mathrsfs}
\usepackage{multirow}
\usepackage{xcolor}
\definecolor{aa}{RGB}{0,0,139}
\usepackage[bookmarksnumbered=true,colorlinks,urlcolor=aa,linkcolor=blue,anchorcolor=aa,citecolor=aa]{hyperref}
\usepackage{gensymb}

\begin{document}
\include{def-com}
\title{Measurements of the branching fractions of \boldmath $\chi_{cJ}\to\Sigma^0\bar{\Sigma}^0\eta$}
\newpage

\date{\today}

\author{
\begin{small}
\begin{center}
M.~Ablikim$^{1}$\BESIIIorcid{0000-0002-3935-619X},
M.~N.~Achasov$^{4,c}$\BESIIIorcid{0000-0002-9400-8622},
P.~Adlarson$^{83}$\BESIIIorcid{0000-0001-6280-3851},
X.~C.~Ai$^{89}$\BESIIIorcid{0000-0003-3856-2415},
C.~S.~Akondi$^{31A,31B}$\BESIIIorcid{0000-0001-6303-5217},
R.~Aliberti$^{39}$\BESIIIorcid{0000-0003-3500-4012},
A.~Amoroso$^{82A,82C}$\BESIIIorcid{0000-0002-3095-8610},
Q.~An$^{78,65,\dagger}$,
Y.~H.~An$^{89}$\BESIIIorcid{0009-0008-3419-0849},
M.~S.~Anderson$^{39}$\BESIIIorcid{0009-0008-1550-2632}
% ...
\\
\vspace{0.2cm}
(BESIII Collaboration)\\
\vspace{0.2cm} {\it
$^{1}$ Institute of High Energy Physics, Beijing 100049, People's Republic of China\\
% ...
$^{4}$ Budker Institute of Nuclear Physics SB RAS (BINP), Novosibirsk 630090, Russia\\
% ...
$^{31}$ INFN Sezione di Ferrara, (A)INFN Sezione di Ferrara, I-44122, Ferrara, Italy; (B)University of Ferrara, I-44122, Ferrara, Italy\\
% ...
$^{39}$ Johannes Gutenberg University of Mainz, Johann-Joachim-Becher-Weg 45, D-55099 Mainz, Germany\\
% ...
$^{65}$ State Key Laboratory of Particle Detection and Electronics, Beijing 100049, Hefei 230026, People's Republic of China\\
% ...
$^{78}$ University of Science and Technology of China, Hefei 230026, People's Republic of China\\
% ...
$^{82}$ University of Turin and INFN, (A)University of Turin, I-10125, Turin, Italy; (B)University of Eastern Piedmont, I-15121, Alessandria, Italy; (C)INFN, I-10125, Turin, Italy\\
$^{83}$ Uppsala University, Box 516, SE-75120 Uppsala, Sweden\\
% ...
$^{89}$ Zhengzhou University, Zhengzhou 450001, People's Republic of China\\
\vspace{0.2cm}
$^{\dagger}$ Deceased\\
% ...
$^{c}$ Also at the Novosibirsk State University, Novosibirsk, 630090, Russia\\
% ...
}
\end{center}
\vspace{0.4cm}
\end{small}
}


\begin{abstract}
Using $(2712.4\pm14.3)\times10^{6}$ $\psi(3686)$ events collected with the BESIII detector at the BEPCII collider, we report the first observations of the decay $\chi_{cJ}(J=0,1,2) \to \Sigma^0\bar{\Sigma}^0\eta$. The signal significances, after considering systematic uncertainties, are determined to be 7.2$\sigma$, 5.7$\sigma$, and 6.7$\sigma$ for $\chi_0$, $\chi_{c1}$, and $\chi_{c2}$, respectively. The branching fractions are measured to be $\mathcal B(\chi_{c0}\to\Sigma^0\bar{\Sigma}^0\eta)=({9.7 \pm 1.7 \pm 0.8}) \times 10^{-5}$,
$\mathcal B(\chi_{c1}\to\Sigma^0\bar{\Sigma}^0\eta)=({4.9 \pm 1.1 \pm 0.4}) \times 10^{-5}$, and
$\mathcal B(\chi_{c2}\to\Sigma^0\bar{\Sigma}^0\eta)=({6.1 \pm 1.2 \pm 0.3}) \times 10^{-5}$, where the first uncertainties are statistical and the second systematic. No significant structure in the $\Sigma^0\bar{\Sigma}^0,\Sigma^0\eta,\bar{\Sigma}^0\eta$ mass distributions is observed under the current statistics.
\end{abstract}

\newcommand{\BESIIIorcid}[1]{\href{https://orcid.org/#1}{\hspace*{0.1em}\raisebox{-0.45ex}{\includegraphics[width=1em]{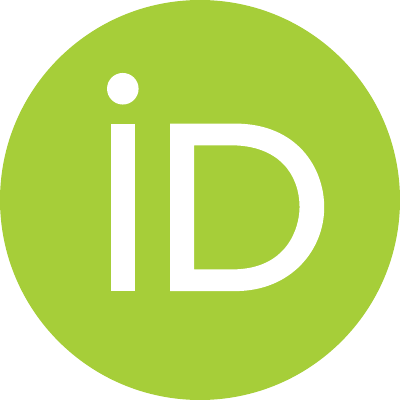}}}}
\maketitle

\section{Introduction}\label{sec:introduction}
Quantum chromodynamics (QCD) has been extensively validated in the high-energy regime. However, in the low-energy domain where perturbative methods are not applicable, numerous unresolved questions and mysteries persist. Charmonium states serve as valuable probes for investigating non-perturbative QCD. The vector charmonium state $\psi$ decays into a baryon pair alongside a pseudo-scalar state are commonly elucidated using  pole models~\cite{Nstarexplain1,Nstarexplain2,Nstarexplain3}. However, a comprehensive understanding of the $\chi_{cJ}$ decays into final state involving baryons remains elusive. Recently, using the large $\psi(3686)$ sample collected by the BESIII detector, along with the radiative transition $\psi(3686) \to \gamma \chi_{cJ}$, we have measured several $\chi_{cJ}$ decays into final states involving baryons, such as $\chi_{cJ} \to \Sigma^0 \bar{\Sigma}^0$ and $\Sigma^+ \bar{\Sigma}^-$~\cite{BESIII:2017szd}, $\Sigma^- \bar{\Sigma}^+$~\cite{BESIII:2020raf}, $ \Lambda \bar{\Lambda} \eta$~\cite{BESIII:2022rsg},  $\Sigma^+ \bar{\Sigma}^- \eta$~\cite{BESIII:2024sfs}, and $\Lambda \bar{\Lambda} \eta'$~\cite{BESIII:2508.18761}. 

To date, the decays $\chi_{cJ} \to \Sigma^0 \bar{\Sigma}^0 \eta$ have not been investigated either experimentally or theoretically. To address this gap, we present the first measurements of these decay channels using a sample of $(2712.4 \pm 14.3) \times 10^6$ $\psi(3686)$ events~\cite{psip} collected with the BESIII detector. We examine the invariant mass spectra of $\Sigma^0\bar{\Sigma}^0$ and $\Sigma^0(\bar{\Sigma}^0)\eta$ systems to search for potential resonant structures. Furthermore, comparisons of the branching fractions for $\chi_{cJ} \to \Sigma^0 \bar{\Sigma}^0 \eta$ and $\chi_{cJ} \to \Sigma^+ \bar{\Sigma}^- \eta$ provide an opportunity to test isospin symmetry in these decays and offer insights into their underlying mechanisms.

\section{BESIII Detector and Monte Carlo Simulation}
\label{sec:simu}
The BESIII detector~\cite{Ablikim:2009aa} records symmetric $e^+e^-$ collisions provided by the BEPCII storage ring~\cite{Yu:IPAC2016-TUYA01} in the center-of-mass energy range from 1.84 to 4.95~GeV, with a peak luminosity of $1.1 \times 10^{33}\;\text{cm}^{-2}\text{s}^{-1}$ achieved at $\sqrt{s} = 3.773\;\text{GeV}$. BESIII has collected large data samples in this energy region~\cite{Ablikim:2019hff,EcmsMea,EventFilter}. The cylindrical core of the BESIII detector covers 93\% of the full solid angle and consists of a helium-based multilayer drift chamber~(MDC), a time-of-flight system~(TOF), and a CsI(Tl) electromagnetic calorimeter~(EMC), which are all enclosed in a superconducting solenoidal magnet providing a 1.0~T magnetic field. The solenoid is supported by an octagonal flux-return yoke with resistive plate counter muon identification modules interleaved with steel. The charged-particle momentum resolution at $1~{\rm GeV}/c$ is $0.5\%$, and specific energy loss ${\rm d}E/{\rm d}x$
resolution is $6\%$ for electrons from Bhabha scattering. The EMC measures photon energies with a
resolution of $2.5\%$ ($5\%$) at $1$~GeV in the barrel (end cap)
region. The time resolution in the plastic scintillator TOF barrel region is 68~ps, while
that in the end cap region was 110~ps. The end cap TOF
system was upgraded in 2015 using multigap resistive plate chamber
technology, providing a time resolution of 60~ps,
which benefits $83\%$ of the data used in this analysis~\cite{etof,etof2,etof3}.

Monte Carlo (MC) simulated data samples produced with a {\sc geant4}-based~\cite{geant4} software package, which includes the geometric description of the BESIII detector and the detector response, are used to determine detection efficiencies and to estimate backgrounds. The simulation models the beam energy spread and initial state radiation (ISR) in the $e^+e^-$ annihilations with the generator {\sc kkmc}~\cite{ref:kkmc1,ref:kkmc2}. The inclusive MC sample includes the production of the $\psi(3686)$ resonance, the ISR production of the $J/\psi$, and the continuum processes incorporated in {\sc kkmc}~\cite{ref:kkmc1,ref:kkmc2}. All particle decays are modelled with {\sc evtgen}~\cite{ref:evtgen,Ping:2008zz} using branching fractions either taken from the Particle Data Group~\cite{pdg}, when available, or otherwise estimated with {\sc lundcharm}~\cite{ref:lundcharm}. The potential background events in the inclusive MC are examined using the event-type analysis tool TopoAna~\cite{TopoAna}. To assess the detection efficiency of each signal decay, $10^6$ signal MC events are generated, following the decay chain $\psi(3686) \to \gamma \chi_{cJ}$, $\chi_{cJ} \to \Sigma^0 \bar{\Sigma}^0 \eta$, $\Sigma^0 \to \gamma \Lambda$, $\bar{\Sigma}^0 \to \gamma \bar{\Lambda}$, $\Lambda \to p \pi^-$, $\bar{\Lambda} \to \bar{p} \pi^+$, and $\eta \to \gamma \gamma$. The decays $\psi(3686) \to \gamma \chi_{cJ}$ are generated using a BESIII specified model, which accounts for the angular distributions of the radiative photons, while the subsequent decays are simulated based on the phase space model.

\section{Event Selection and Background Analysis}\label{sec:selection}

Signal candidates are reconstructed through their complete decay chains (yielding the $\pi^+\pi^-p\bar{p}5\gamma$ final states), and candidate events are processed as outlined below. $\bar{\Lambda}$ and $\Lambda$ are reconstructed by utilizing $\pi^+\bar{p}$ and $\pi^- p$ pairs, respectively. Additionally, $\Sigma^0(\bar{\Sigma}^0)$ are reconstructed by combining $\Lambda(\bar{\Lambda})$ with selected photons. The $\eta$ meson is reconstructed using a pair of photons. The remaining photon is thus designated as the radiative photon.

Charged tracks detected in the MDC must be confined to an angular region of $|\cos\theta| < 0.93$, with the polar angle $\theta$ defined relative to the symmetry axis of the MDC. Given the relatively long lifetimes of $\Lambda$ , no specific vertex requirement is imposed on the charged tracks from the $\Lambda$ decays. Each candidate event is expected to include a minimum of two positive charged tracks and two negative charged tracks.

No particle identification (PID) is applied to the charged tracks. Instead, particle types are established by combining positive and negative tracks with  various particle type hypotheses. Following a vertex fit to the candidate $p\pi$ pairs, only combinations yielding the minimum value of $\sqrt {{{\left( {{M_{p{\pi ^-}}} - m_{\Lambda }} \right)}^2} + {{\left( {{M_{\bar p{\pi ^+}}} - {m_{\bar {\Lambda} } }} \right)}^2}}$ are retained for further analysis. Here, $m_\Lambda$ and $m_{\bar{\Lambda}}$ denote the world-averaged masses of $\Lambda$ and $\bar{\Lambda}$~\cite{pdg}, the $M_{p{\pi ^-}}$ and $M_{p{\pi ^+}}$ denote the invariant masses of the $p{\pi ^-}$ and $p{\pi ^+}$ combinations, respectively. Additionally, we require the decay length of $\Lambda / \bar{\Lambda}$ to exceed zero, and the invariant mass of $p\pi$ to be within a $\pm$9 MeV/$c^{2}$ range around the nominal mass of $\Lambda$.

Photon candidates are identified using isolated showers in the EMC.  The deposited energy of each shower must be more than 25~MeV in the barrel region ($|\cos \theta|< 0.80$) and more than 50~MeV in the end cap region ($0.86 <|\cos \theta|< 0.92$). 
To exclude showers that originate from
charged tracks,
the angle subtended by the EMC shower and the position of the closest charged track at the EMC
must be greater than 10 degrees as measured from the IP; for $\bar{p}$ tracks, the opening angle greater than 20 degrees
To suppress electronic noise and showers unrelated to the event, the difference between the EMC time and the event start time is required to be within [0, 700]\,ns. The total number of photons must be not less than five.

A four constraints (4C) kinematic fit is applied to each $\psi(3686)$ candidate: this fit constrains the four-momentum of the final-state system ($p\bar{p}\pi^+\pi^-5\gamma$) to match that of the initial $e^+e^-$ collision system, for the purpose of enhancing resolution and reducing the background level. When the number of photons exceeds five, all possible combinations are tested, and the combination yielding the minimum $\chi^2_{4\rm C}$ is retained for further analysis. A selection cut on the $\chi_{4\rm C}^{2}$ value from the fit is applied. The selection criterion is chosen by optimizing the $\chi_{4\rm C}^{2}$ Figure of Merit (FOM) value $S/\sqrt{S+B}$, where $S$ and $B$ represent the numbers of signal events from  the signal MC sample and the number of background events from the inclusive MC sample, respectively. The signal yield $S$ has been normalized based on the preliminary results obtained in this analysis. Following optimization, candidate events are required to satisfy $\chi^2_{4\rm C}<20$. To suppress backgrounds with one less or more photon in the final states, we require $\chi_{5\gamma}^2$ being less than either $\chi_{4\gamma}^2$ or $\chi_{6\gamma}^2$, where the value of $n$ in $\chi_{n\gamma}^2$ corresponds to the number of photons considered in the 4C kinematic fit. This selection criterion suppresses approximately $50\%$ of the background contributions while incurring only about $5\%$ loss in signal efficiency.

\begin{figure*}[htbp]
\begin{center}
\includegraphics[width=0.3\textwidth]{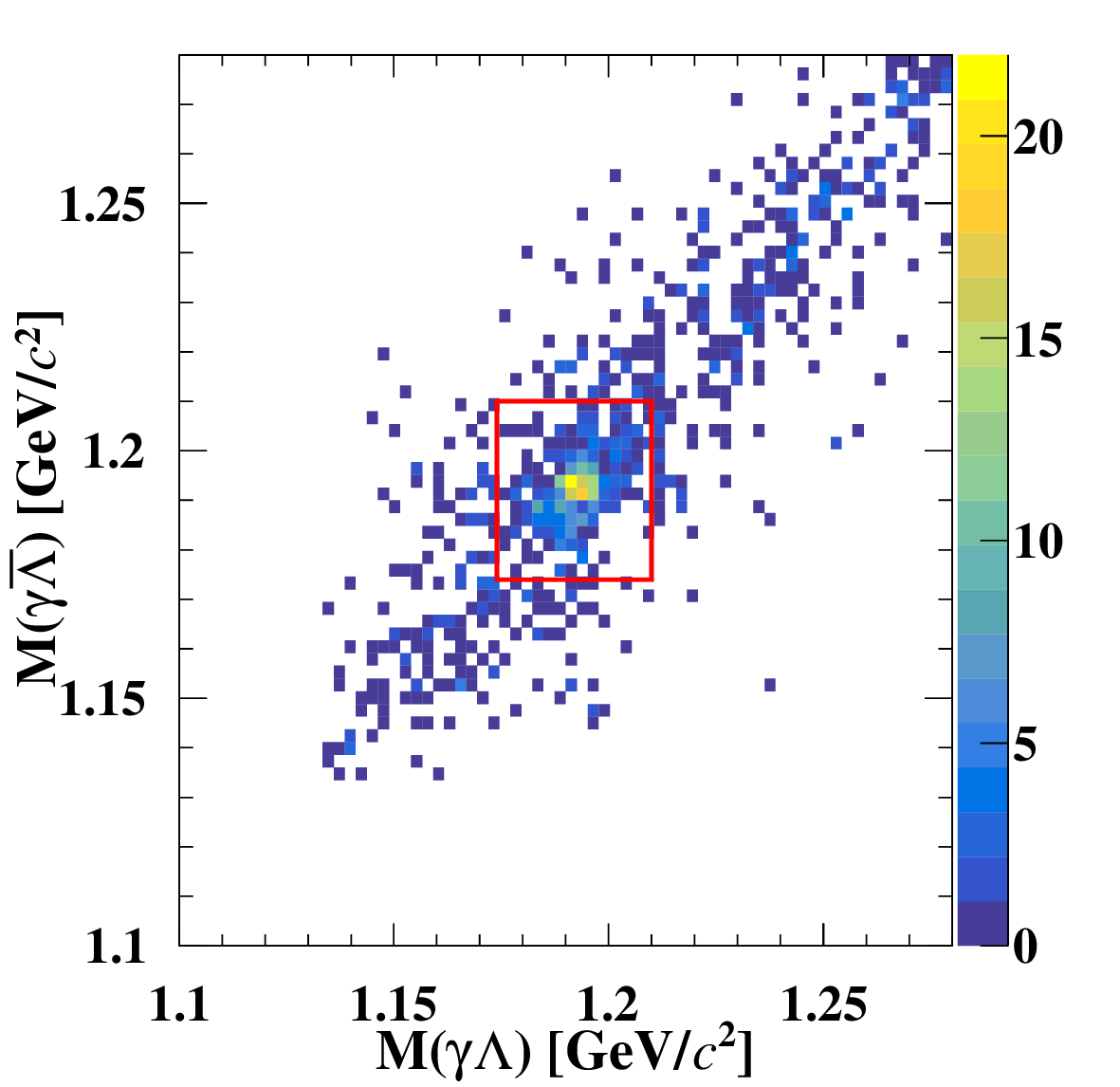}
\includegraphics[width=0.3\textwidth]{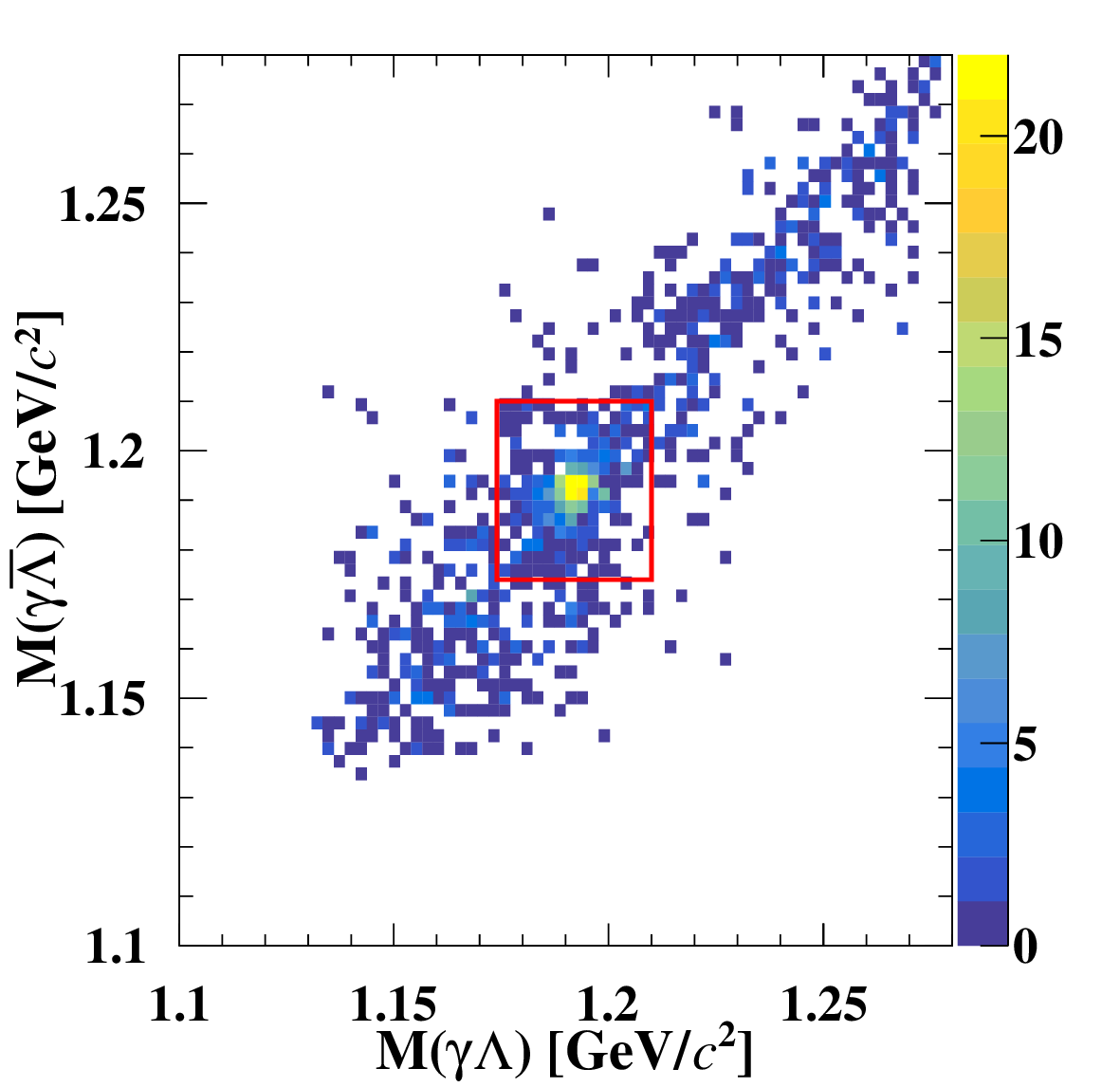}
\includegraphics[width=0.3\textwidth]{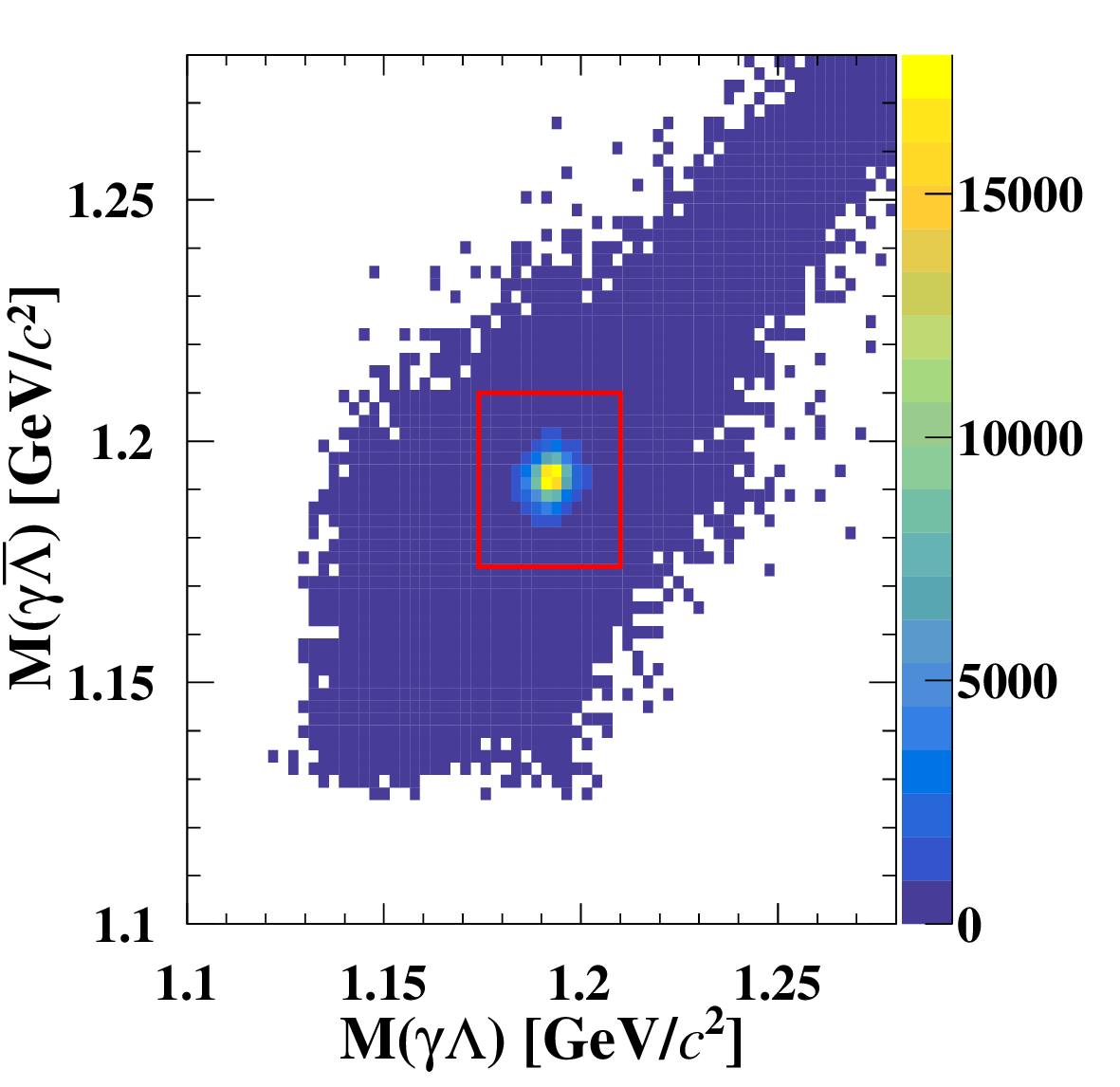}
\caption{The distributions of $M(\gamma\Lambda)$ versus $M(\gamma\bar{\Lambda}))$ of (left) data, (middle) inclusive MC, and (right) signal MC samples. The red squares indicate the $\Sigma^0\bar \Sigma^0$ signal regions.}
\label{fig:gammaLambda}
\end{center}
\end{figure*}

The hyperons $\Sigma^0$ and $\bar{\Sigma}^0$ are reconstructed using $\gamma\Lambda$ and $\gamma\bar{\Lambda}$ pairs, respectively. From the five previously determined photons in the kinematic fit, the two corresponding photons are selected by minimizing ${{| {{M_{\gamma_1\Lambda}} - M_{\gamma_2\bar{\Lambda} }} |}} $, where $\gamma_1$ and $\gamma_2$ indicate any two possible photons selected from the five. Here, $M_{\gamma{\Lambda}}$ denotes the invariant mass of the $\gamma\Lambda$ combination. Additionally, they are required to fall within $\pm18$ MeV/$c^{2}$ of the nominal mass $m_{\Sigma^0}$. The distributions of $M_{\gamma\Lambda}$ versus $M_{\gamma\bar{\Lambda}}$ from data and MC samples are illustrated in Fig.~\ref{fig:gammaLambda}, revealing distinct $\Sigma^0/\bar{\Sigma}^0$ pairs. 

The $\eta$ meson is reconstructed from the three remaining photons, with all possible photon pair combinations considered as $\eta$ candidates. The signal region for $\eta$ is defined as a $\pm27$ MeV/$c^{2}$ window around its nominal mass, as depicted in Fig.~\ref{scalefactor}, while the sideband regions are set as ($439$, $493$) MeV/$c^{2}$ and ($601$, $655$) MeV/$c^{2}$. Events in the sideband regions are utilized to estimate the combinatorial $\gamma\gamma$ contributions.

To suppress backgrounds from $\psi(3686)\to \gamma \gamma J/\psi$, the invariant mass of the recoil system
  for any photon pair must fall outside the $J/\psi$ mass window. To suppress backgrounds involving final states containing $\pi^0$, the invariant mass of the radiative photon $\gamma^{E1}$ and any other $\gamma$ must be outside the $\pi^0$ mass window. All the aforementioned vetoing mass windows are optimized using the FOM approach and listed in Table~\ref{tab:veto}.

\begin{table}
    \centering
   \caption{Veto ranges for various photon combinations. Here, Rec and Inv indicate recoil mass and invariant mass, respectively. Meanwhile, $\gamma^\eta_{H/L}$, $\gamma^{E1}$, and $\gamma^{\Sigma^0/\bar \Sigma^0}$ represent the Higher/Lower energy photons from $\eta$ decays, the radiative photon from $\psi(3686)$, and the radiative photons from $\Sigma^0/\bar \Sigma^0$ decays, respectively.}
    \label{tab:veto}
    \begin{tabular}{cc}
    \\ \hline \hline
      Mass system   & Window width (MeV/$c^2$) \\
      \hline
      $|\mathrm{Rec}(\gamma^\eta_H \gamma^\eta_L) - m_{J/\psi}|$   & 17.0 \\
      $|\mathrm{Rec}(\gamma^\eta_H \gamma^{E1}) - m_{J/\psi}|$  & 8.0 \\
      $|\mathrm{Inv}(\gamma^\eta_H \gamma^{E1}) - m_{\pi^0}|$   &  17.0 \\
      $|\mathrm{Inv}(\gamma^\eta_L \gamma^{E1}) - m_{\pi^0}|$   &  5.0 \\
      $|\mathrm{Inv}(\gamma^{\Sigma^0} \gamma^{E1}) - m_{\pi^0}|$   &  5.0 \\
      $|\mathrm{Inv}(\gamma^{\bar{\Sigma}^0} \gamma^{E1}) - m_{\pi^0}|$   &  7.0 \\
      \hline \hline
    \end{tabular}
\end{table}

After applying all the aforementioned selection criteria, the analysis of the inclusive MC samples indicates that the remaining $\chi_{cJ}$ peaking backgrounds do not involve the $\eta$ meson as an intermediate state. Therefore, we utilize events in the $\eta$ sideband regions to characterize the peaking backgrounds in the subsequent signal yield determination.

\section{Signal yield and Branching fractions}\label{Sec:sig}
Following the methodology applied in Ref.~\cite{Ablikim:2017wt}, the signal yields for the three decay channels \(\chi_{cJ} \to \Sigma^0 \bar{\Sigma}^0 \eta\) are determined by simultaneously fitting the \(M_{\Sigma^0 \bar{\Sigma}^0 \gamma \gamma}\) distributions of events in both the \(\eta\) signal and sideband regions.

For events in the \(\eta\) signal region, the fit function comprises three components: the signal, peaking background, and smooth background, expressed as
\[
(N^{S} + f_\eta N_{\eta \text{sb}}^{S}) F^{S} + N^{B} F^{B} \ .
\]
For events in the \(\eta\) sideband region, the fit function consists of the signal and background components:
\[
N^{S}_{\eta \text{sb}} F^{S} + N^{B}_{\eta \text{sb}} F^{B}_{\eta \text{sb}} \ .
\]
Here, \(N^{S}\) (\(N^{S}_{\eta \text{sb}}\)) and \(N^{B}\) (\(N^{B}_{\eta \text{sb}}\)) represent the numbers of $\chi_{cJ}$ signal and background events in the \(\eta\) signal (sideband) region, respectively, as determined through the fit. The function $F^S$ describes the $\chi_{cJ}$ signal shape and is shared between the fits to both the $\eta$ signal and sideband regions.  It is defined as
\[
F^{S} = (E^{3}_{\gamma^{E1}} \times \text{BW}) \otimes \text{DGaus} \ .
\]
Here, \(E_{\gamma^{E1}}\) is the energy of the radiative photon whose third power is required to describe the E1 transition~\cite{E3gammaFunction}, \(\text{BW}\) is the non-relativistic Breit-Wigner function for the \(\chi_{cJ}\) resonance, and \(\text{DGaus}\) is a double Gaussian function accounting for the detector resolution. The parameters of the \(\text{BW}\) function are fixed to the world-average values from the PDG~\cite{pdg}, while those of the \(\text{DGaus}\) function are determined using signal MC samples and are fixed during the fit.

\begin{figure}[htbp]
\begin{center}
\includegraphics[width=0.4\textwidth]{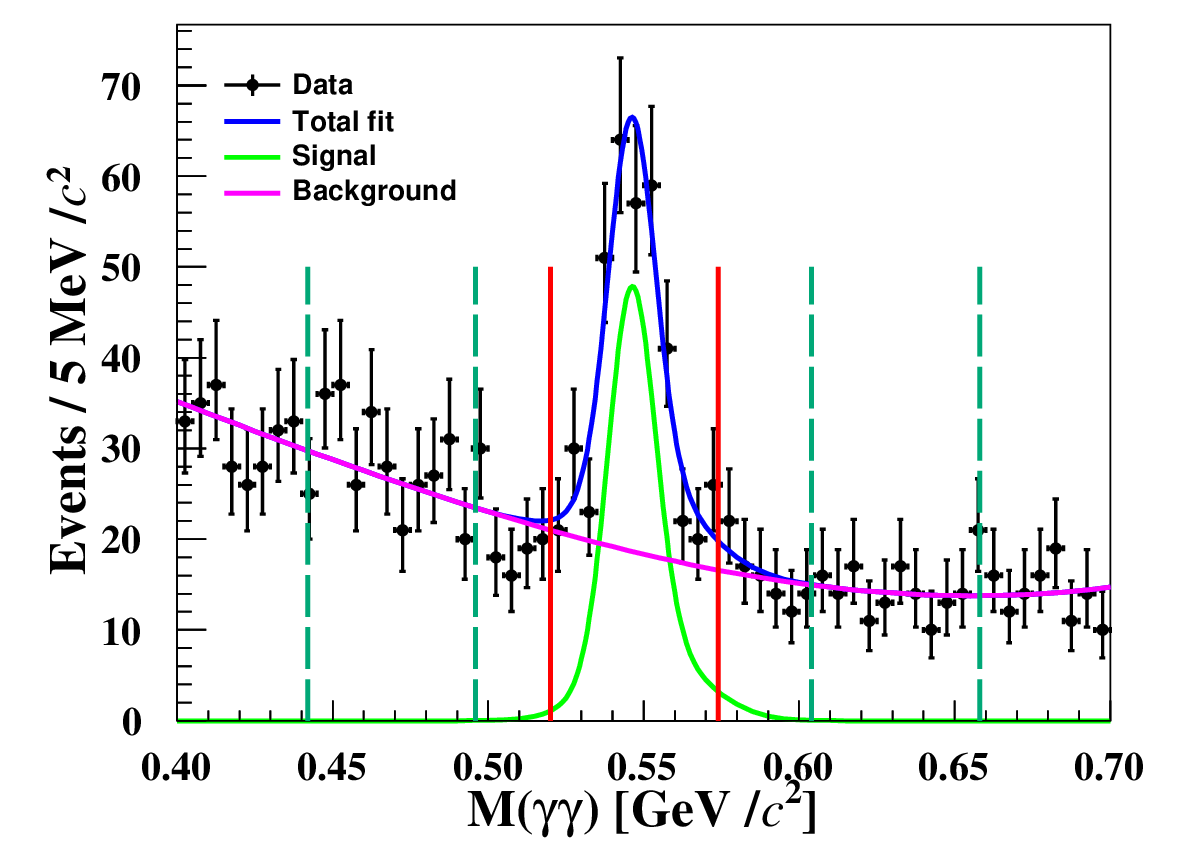}
\caption{Distribution of all $M(\gamma\gamma)$ combinations. The solid blue, solid green, and pink curves correspond to the total, signal, and background contributions, respectively. The vertical solid red and dashed green lines indicate the $\eta$ signal and sideband regions, respectively.}
\label{scalefactor}
\end{center}
\end{figure}

The normalization factor \(f_\eta\) reflects the ratio of non-\(\eta\) backgrounds between the \(\eta\) signal and sideband regions. It is determined to be $0.46 \pm 0.02$ by fitting the invariant mass (\(M_{\gamma\gamma}\)) distribution, as shown in Fig.~\ref{scalefactor}, and integrating the background contributions in the respective regions. During the fit to extract the $\chi_{cJ}$ yields, \(f_\eta\) is fixed at 0.46. Functions \(F^{B}\) and \(F^{B}_{\eta \text{sb}}\) model the non-peaking background shapes in the fits of the \(\eta\) signal and sideband regions, respectively, and both are described by three order Chebyshev polynomial functions with free parameters.

\begin{figure*}[htbp]
\begin{center}
\includegraphics[width=0.8\textwidth]{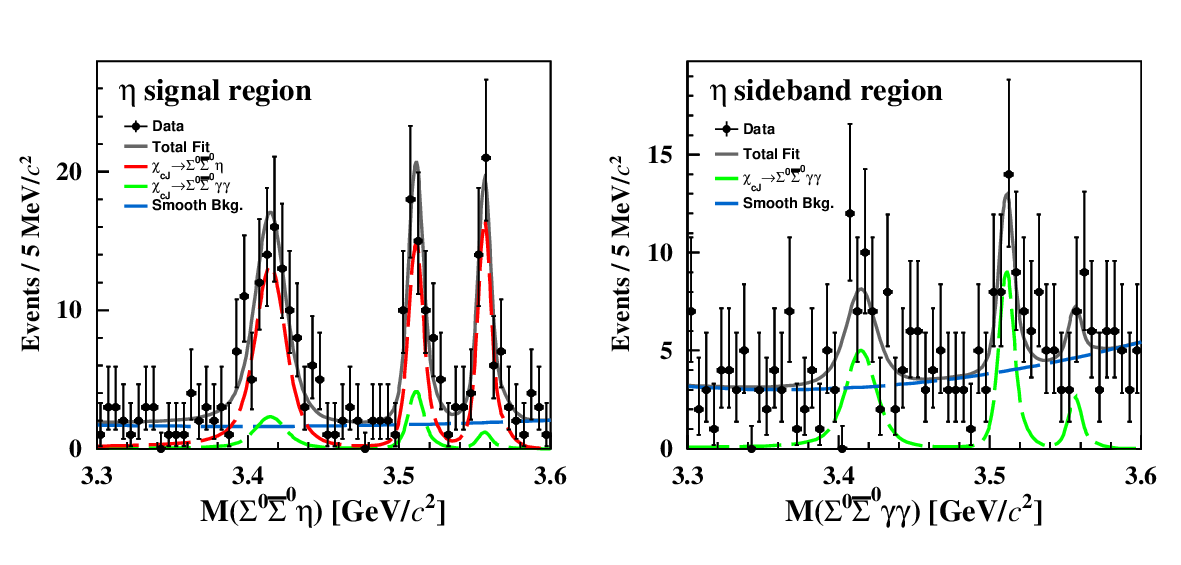}
\caption{Simultaneous fit to the $M_{\Sigma^0\bar{\Sigma}^0\eta}$ distributions in the $\eta$ signal (left) and sideband (right) regions. In both regions, the solid gray curves represent the total contributions. In the $\eta$ signal region, the dashed red, green, and blue lines denote the contributions of the signal, peaking background, and smooth background, respectively. In the $\eta$ sideband regions, the dashed green and blue lines represent the contributions of the $\chi_{cJ}$ signal and background, respectively.}
\label{fit_result}
\end{center}
\end{figure*}

\begin{figure*}[htbp]
\begin{center}
\begin{minipage}[t]{0.3\linewidth}
\includegraphics[width=1\textwidth]{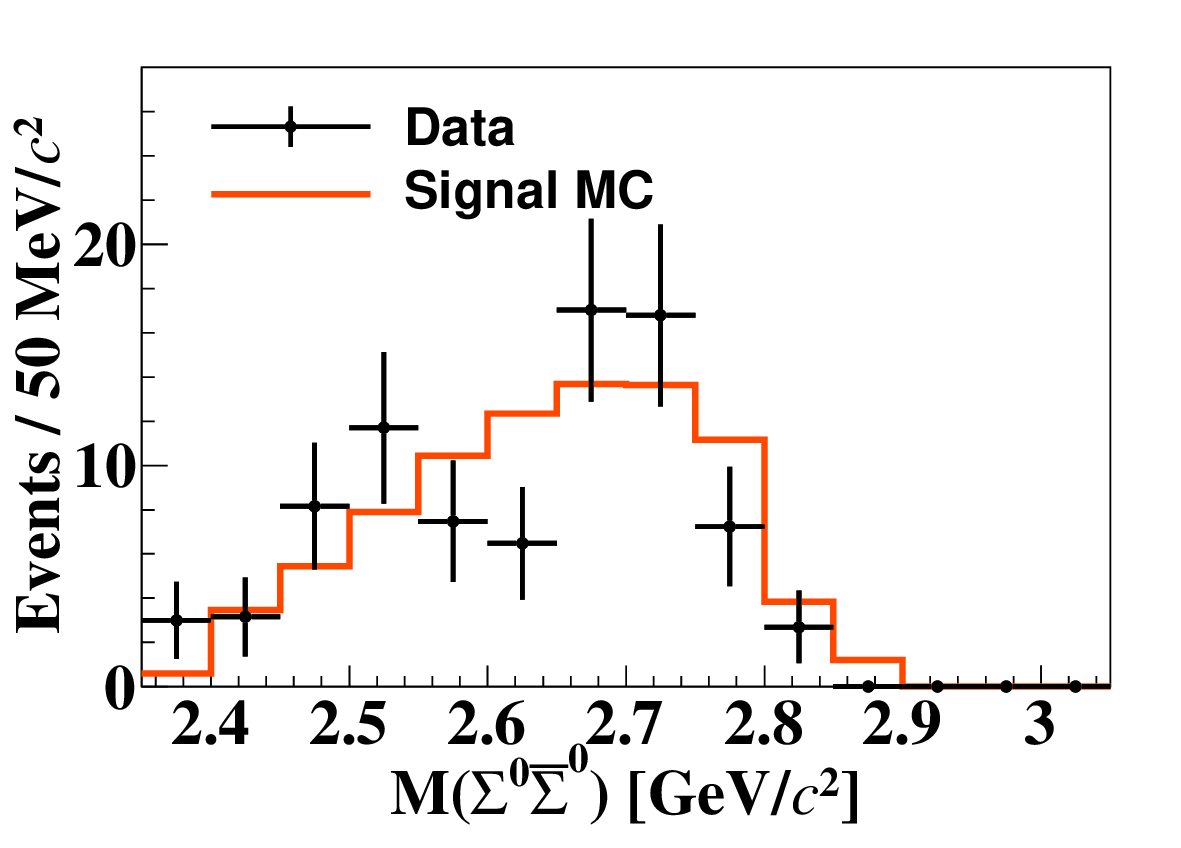}
\end{minipage}
\begin{minipage}[t]{0.3\linewidth}
\includegraphics[width=1\textwidth]{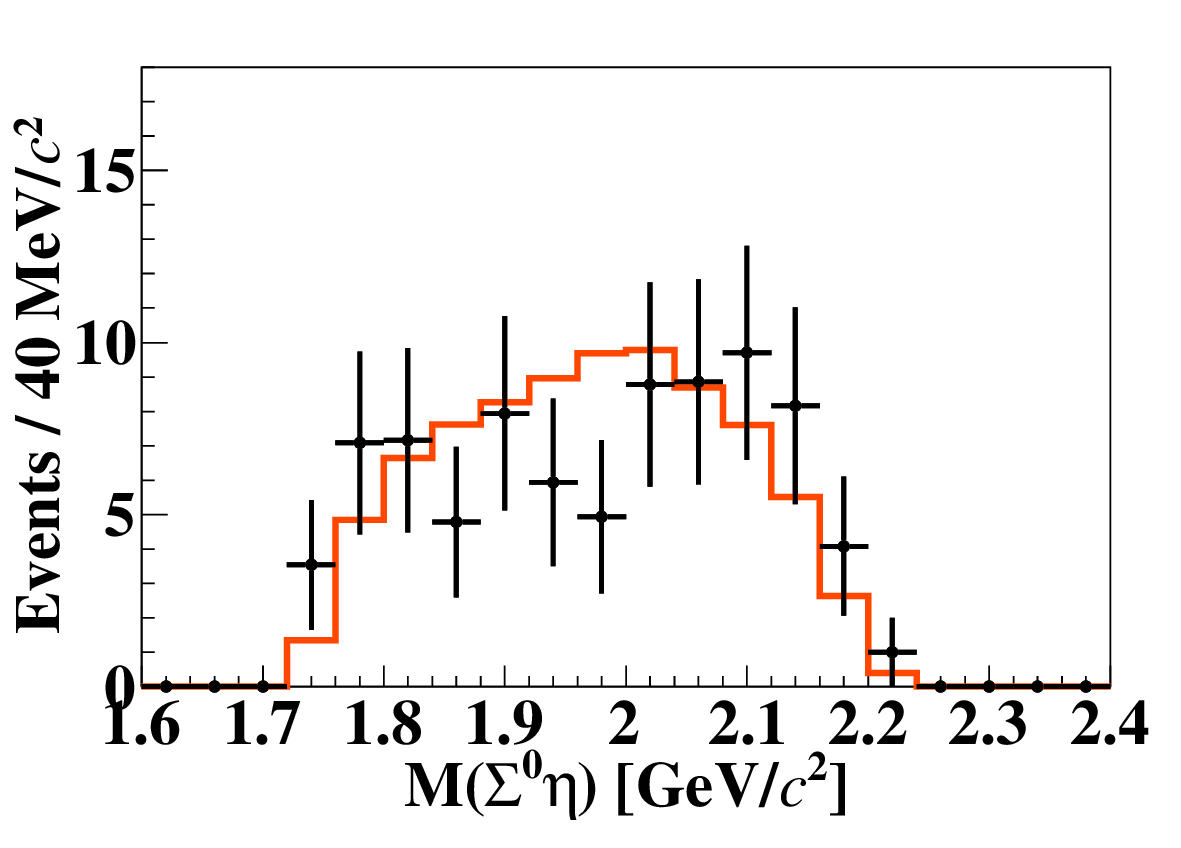}
\end{minipage}
\begin{minipage}[t]{0.3\linewidth}
\includegraphics[width=1\textwidth]{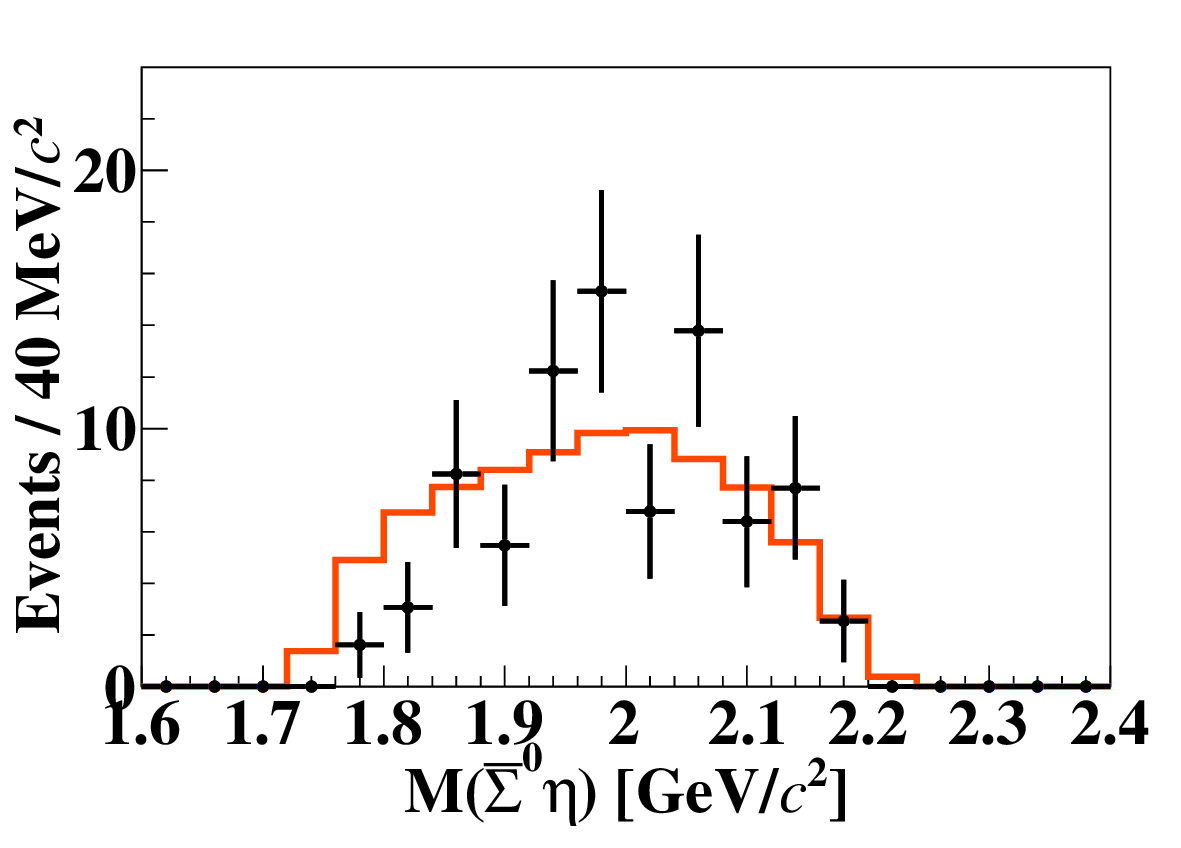}
\end{minipage}

\begin{minipage}[t]{0.3\linewidth}
\includegraphics[width=1\textwidth]{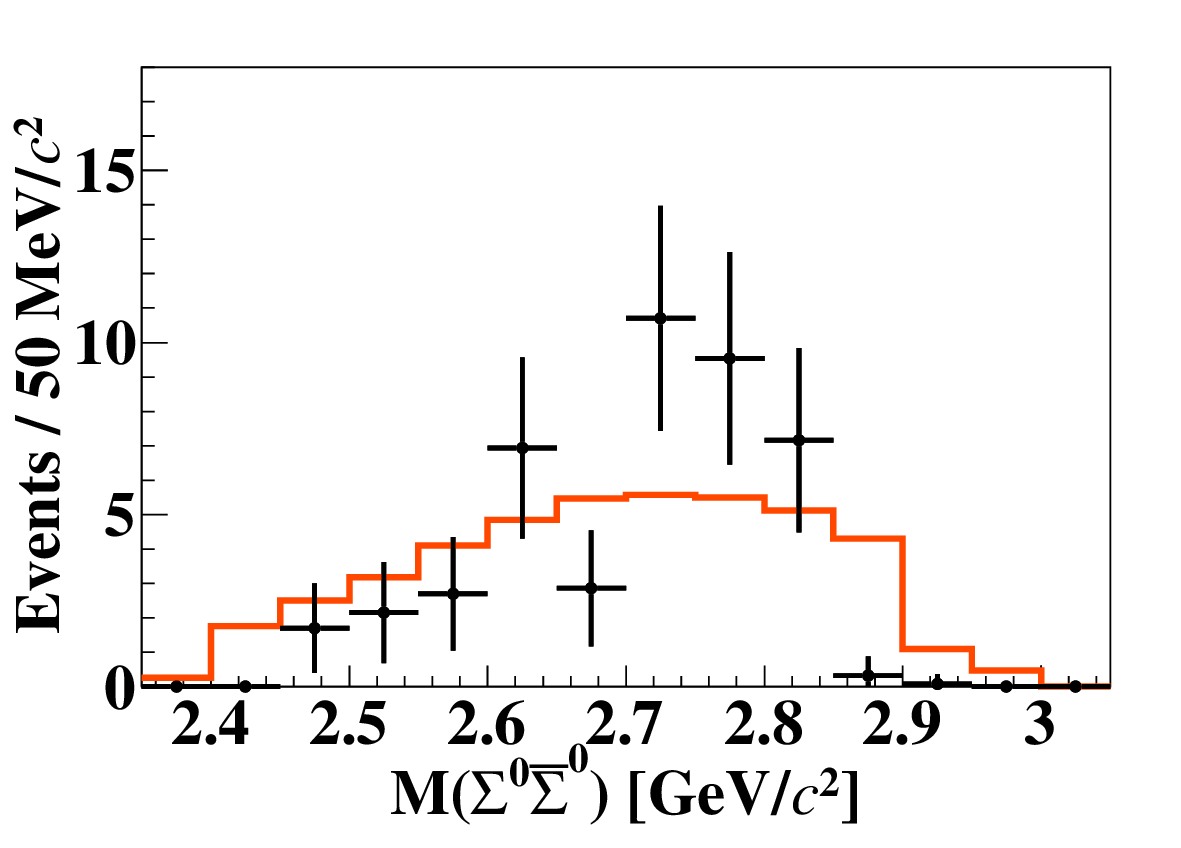}
\end{minipage}
\begin{minipage}[t]{0.3\linewidth}
\includegraphics[width=1\textwidth]{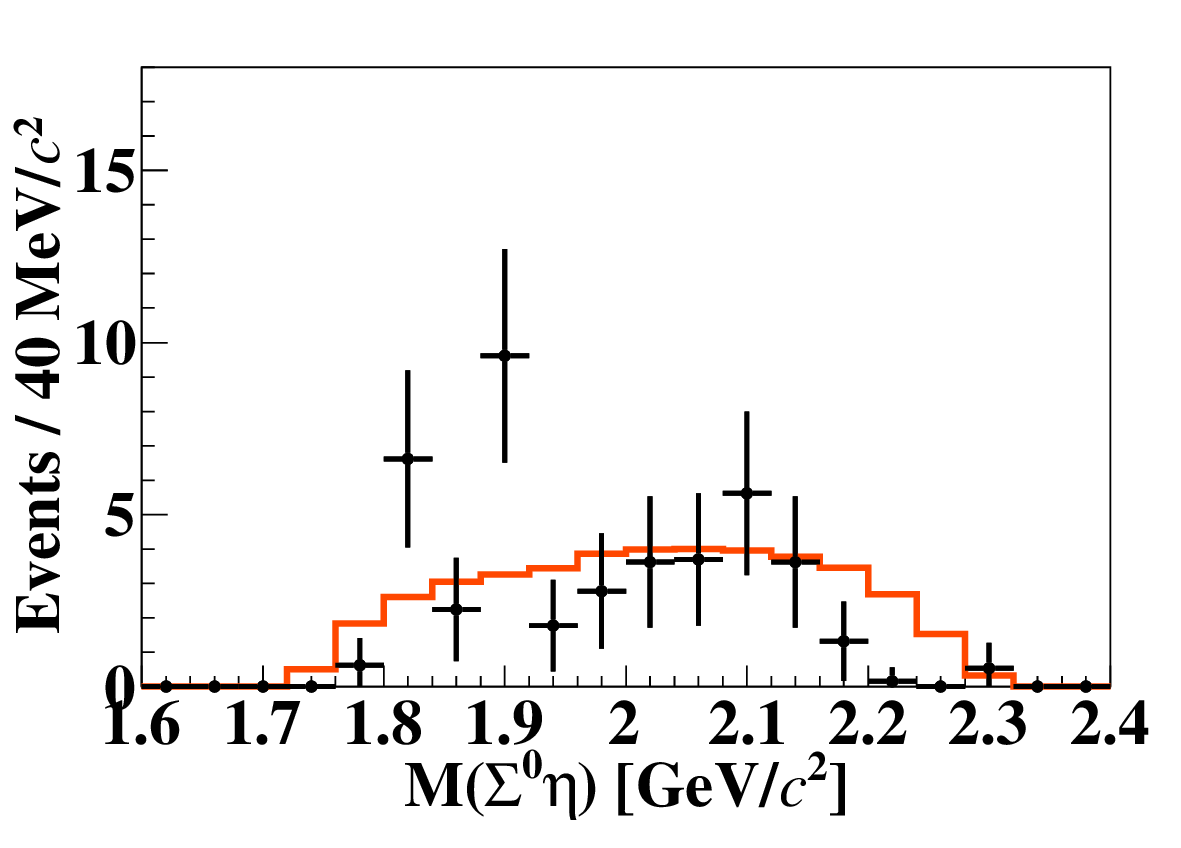}
\end{minipage}
\begin{minipage}[t]{0.3\linewidth}
\includegraphics[width=1\textwidth]{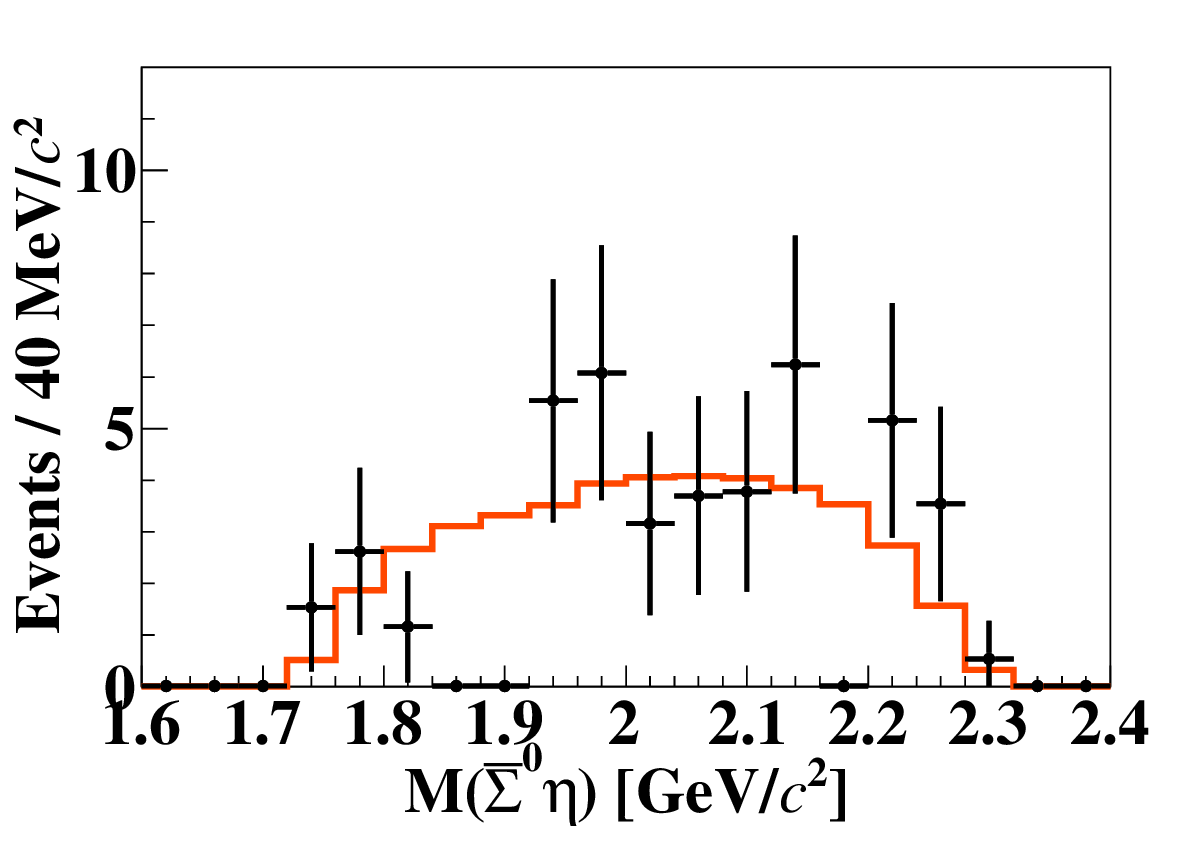}
\end{minipage}

\begin{minipage}[t]{0.3\linewidth}
\includegraphics[width=1\textwidth]{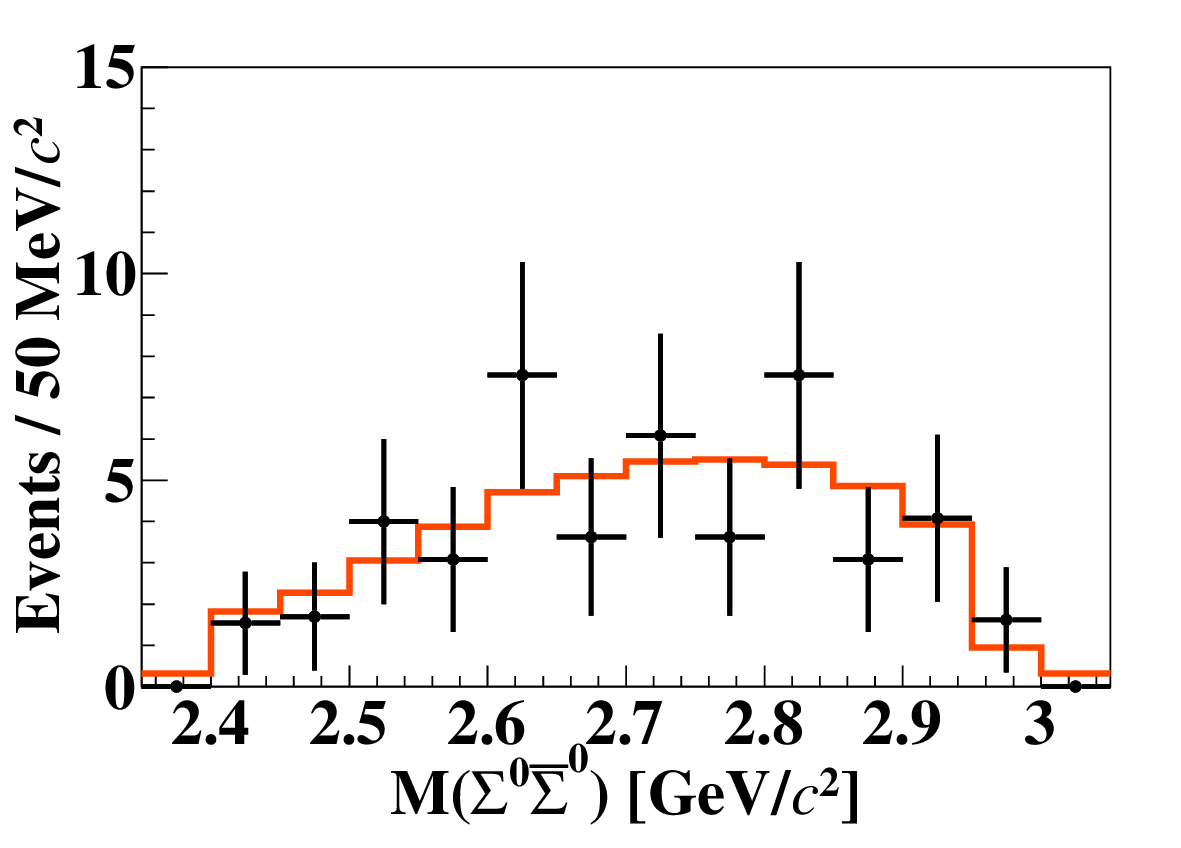}
\end{minipage}
\begin{minipage}[t]{0.3\linewidth}
\includegraphics[width=1\textwidth]{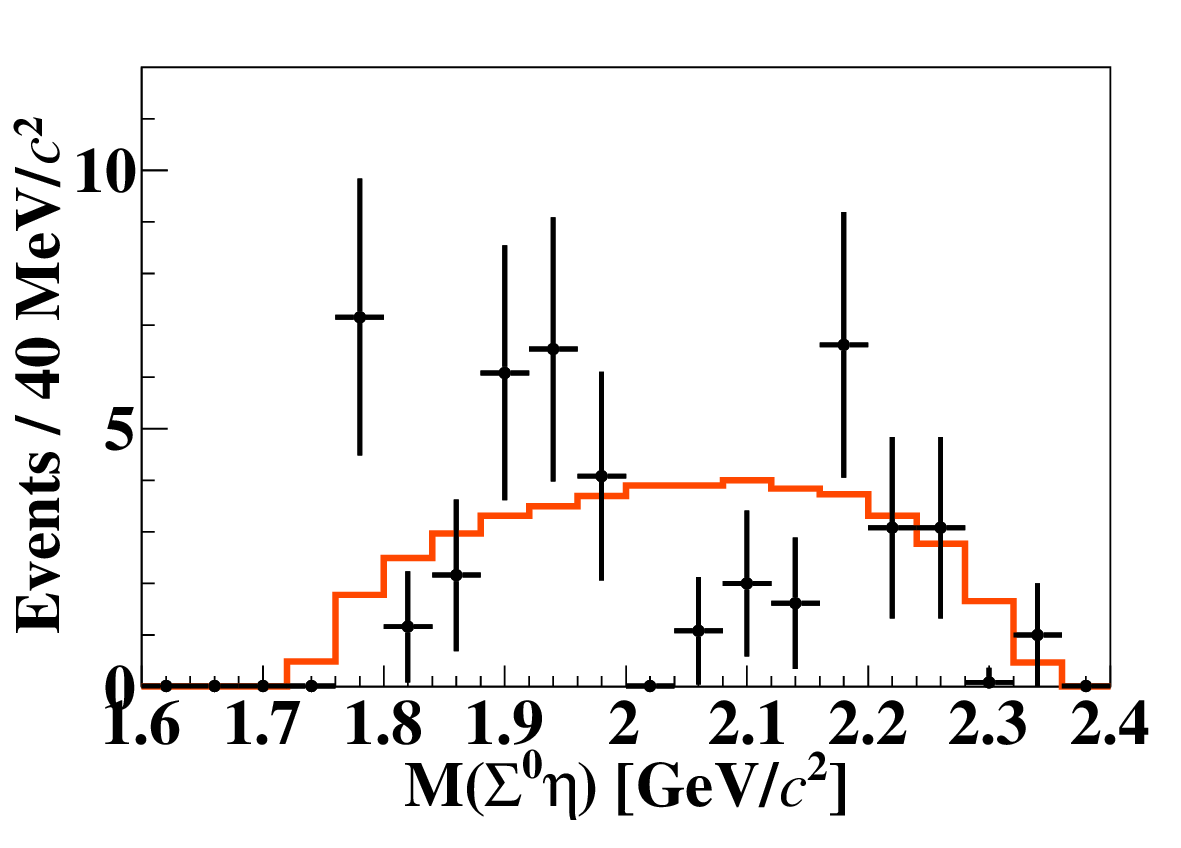}
\end{minipage}
\begin{minipage}[t]{0.3\linewidth}
\includegraphics[width=1\textwidth]{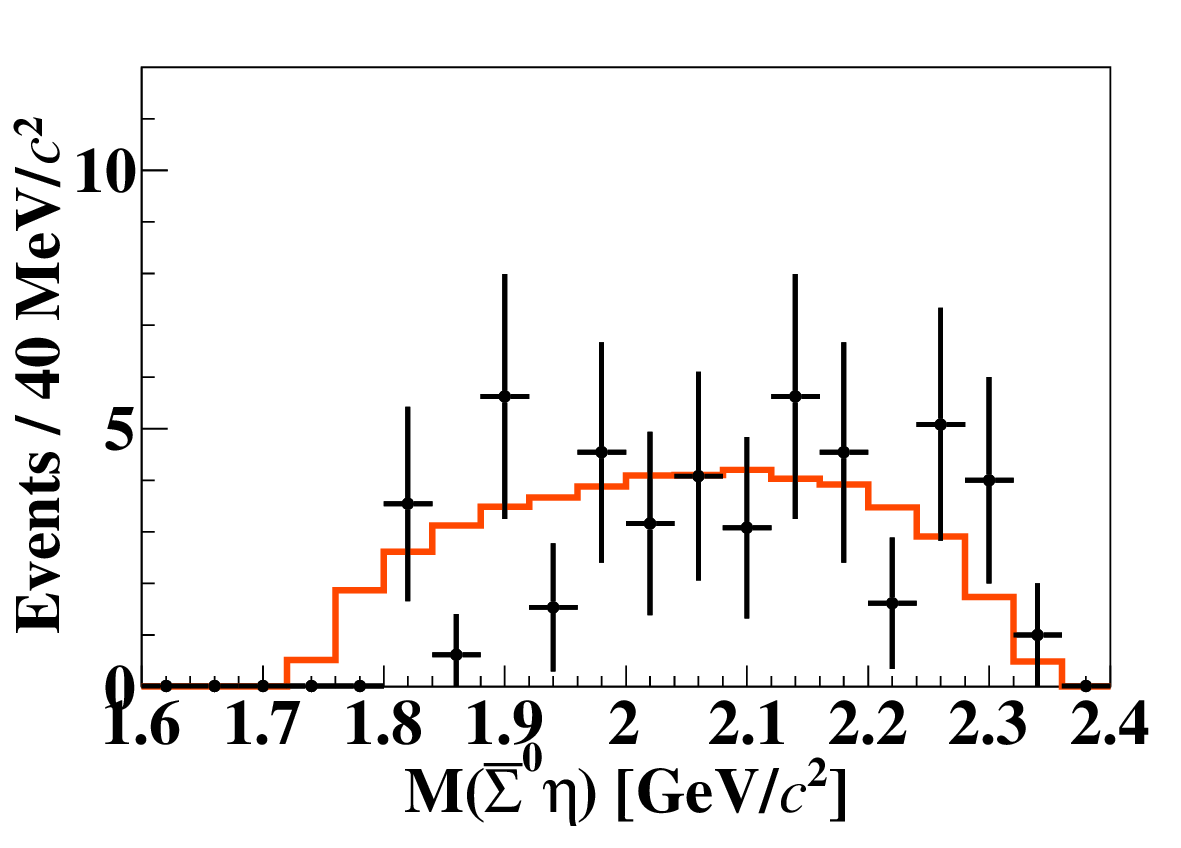}
\end{minipage}
\caption{Invariant mass distributions of various two-body combinations of $\chi_{c0} \to \Sigma^0 \bar{\Sigma}^0 \eta$ (top row), $\chi_{c1} \to \Sigma^0 \bar{\Sigma}^0 \eta$ (middle row), and $\chi_{c2} \to \Sigma^0 \bar{\Sigma}^0 \eta$ (bottom row). Events are displayed after subtracting backgrounds with the data events in the $\eta$ sideband. The signal regions of $\chi_{\rm cJ}$ are defined as $(3.375,3.453)~\rm GeV/c^2$, $(3.489,3.531)~\rm GeV/c^2$, and $(3.535,3.577)~\rm GeV/c^2$ for $\chi_{c0}$, $\chi_{c1}$, and $\chi_{c2}$, respectively.}
\label{newFig:hist_mLP_m2lamb_chic1}
\end{center}
\end{figure*}

The results of the simultaneous fit are presented in Fig.~\ref{fit_result}, with the fitted yields detailed in Table~\ref{eff}. The signal significances of the $\chi_{cJ} \to \Sigma^0 \bar{\Sigma}^0 \eta$ decays are determined to be 7.2$\sigma$, 5.7$\sigma$, and 6.7$\sigma$, respectively, by evaluating the difference in likelihoods with or without inclusion of the signal component in the fit, as well as considering~\cite{Ding:2023cqo} the systematic uncertainties discussed later. Potential intermediate states involving the $\Sigma^0$, $\bar{\Sigma}^0$, and $\eta$ particles are explored by examining the invariant mass distributions of any two-body combinations among the three, as shown in Fig.~\ref{newFig:hist_mLP_m2lamb_chic1}. No discernible structures are evident from these distributions.

\begin{table*}[htbp]
\caption{The fitted numbers of events in the $\eta$ signal and sideband regions $(N^S, N^S_{\eta sb})$, scale factors $(f_\eta)$, detection efficiencies $(\epsilon)$, and the product of the branching fractions $(\mathcal B)$.}
\begin{center}

\setlength{\tabcolsep}{4.5mm} %
\begin{tabular}{c  c  c  c}
\hline \hline
 Channel  &  $\chi_{\rm c0}$  &  $\chi_{\rm c1}$  & $\chi_{\rm c2}$   \\  \hline

  ${N}^{S}$                                & $80.2\pm13.5$    & $43.1\pm9.5$    &  $45.3\pm8.9$ \\ 
  ${N}^{S}_{\eta sb}$      & $30.5\pm9.4$     & $26.3\pm7.6$     &  $7.2\pm6.2$ \\
  $f_{\eta}$  & $0.46$             & $0.46$             &  $0.46$ \\  
  $\epsilon(\%)$                                    & $1.87$            & $2.05$            &  $1.76$ \\
  $\mathcal{B}(\psi(3686)\to\gamma\chi_{\rm c J})\cdot\mathcal{B}(\chi_{\rm cJ} \to \Sigma^0\bar{\Sigma}^0\eta)(\times 10^{6})$ & $9.8\pm1.7\pm0.5$ & $4.8\pm1.1\pm0.3$ & $5.9\pm1.1\pm0.3$\\
  $\mathcal{B}(\chi_{\rm cJ} \to \Sigma^0\bar{\Sigma}^0\eta)(\times 10^{5})$ & $9.7\pm1.7\pm0.6$ & $4.9\pm1.1\pm0.3$ & $6.1\pm1.2\pm0.4$\\
  \hline   \hline
\end{tabular}
\end{center}
\label{eff}
\end{table*}

The branching fractions of $\chi_{cJ}\to\Sigma^0\bar{\Sigma}^0\eta$ are calculated as 
\begin{equation}
{\mathcal B}({\chi_{cJ}} \to \Sigma^0\bar{\Sigma}^0\eta ) = \frac{N^S} {N_{\psi(3686)}\, \epsilon \prod\limits_{i}  {\mathcal B_i} } \ ,
\end{equation}
where $N_{\psi(3686)} = (2712.4\pm14.3)\times10^{6}$~\cite{Ablikim:2017wyh,psip} is the total number of $\psi(3686)$ events in data, $\epsilon$ is the detection efficiency, and $\prod _i {\mathcal B_i}$ is the product of branching fractions, including $\mathcal B(\psi(3686) \rightarrow \gamma\chi_{cJ})$, $\mathcal B(\Sigma^0 \rightarrow \Lambda\gamma)$, $\mathcal B(\bar{\Sigma}^0 \rightarrow \bar{\Lambda}\gamma)$, $\mathcal B(\Lambda \rightarrow p\pi^-)$, $\mathcal B(\bar{\Lambda} \rightarrow \bar{p}\pi^+)$, and $\mathcal B(\eta \rightarrow \gamma \gamma)$~\cite{pdg}. The fitted numbers of events in the $\eta$ signal and sideband regions, scale factor, and detection efficiencies ($\epsilon$) and the product of the branching fractions are listed in Table~\ref{eff}. Then, the branching fractions are determined to be $\mathcal{B}(\chi_{\rm c0}\to \Sigma^0 \bar{\Sigma}^0 \eta)= (9.7 \pm 1.7)\times 10^{-5}$, $\mathcal{B}(\chi_{\rm c1}\to \Sigma^0 \bar{\Sigma}^0 \eta)= (4.9 \pm 1.1)\times 10^{-5}$, $\mathcal{B}(\chi_{\rm c2}\to \Sigma^0 \bar{\Sigma}^0 \eta)= (6.1 \pm 1.2) \times 10^{-5}$. Here, only the statistical uncertainties are presented.

\section{Systematic Uncertainties}
Systematic uncertainties associated with the branching fractions stem from the reconstruction of \(\Lambda(\bar{\Lambda})\) and photons, mass window definitions for resonance signals and veto backgrounds, kinematic fit, signal yield extraction, physics models, quoted branching fractions, and the total number of \(\psi(3686)\) events.

The systematic uncertainty from $\Lambda (\bar{\Lambda})$ reconstruction is studied with the control sample of $J/\psi \to pK^{-} \bar{\Sigma}^{0}(\gamma \bar{\Lambda})+ c.c.$~\cite{Lam_antiLam_phi}. The estimated uncertainties are $1.8\%$ for $\Lambda$ and $1.5\%$ for $\bar{\Lambda}$.

The systematic uncertainty associated with photon reconstruction is estimated to be 0.5$\%$ for each photon with the control sample of  $J/\psi\to\pi^+\pi^-\pi^0$~\cite{photon_uncer}. Given the presence of five photons in each event candidate, the total systematic uncertainty is $2.5\%$.

The uncertainties of the signal regions of $\Sigma^0$, $\Lambda$, and $\eta$ are evaluated using the control samples of  $\psi(3686) \to \eta J/\psi$ and $J/\psi \to \Sigma^0 \bar{\Sigma}^0 \to \gamma\gamma \Lambda\bar{\Lambda}$. The differences in the acceptance efficiencies between data and MC simulation are found negligible.

Uncertainties related to the veto for recoil mass and invariant mass are evaluated via the Barlow test~\cite{Barlow}. The veto mass window is expanded or contracted in five steps of 1 MeV each; for $(\gamma^\eta_L \gamma^{E1})$ and $(\gamma^{\Sigma^0}\gamma^{E1})$, however, the maximum variation of the window range is set to 0.5 MeV. The deviations of the measured branching fractions across all window adjustments are less than $(2\sigma)$, indicating that the corresponding uncertainty is negligible.

The uncertainty associated with the 4C kinematic fit is estimated by comparing the efficiency with and without correction applied to track helix parameters, following the method described in Ref.~\cite{BESIII:2012mpj}.  They are assigned as $1.8\%$, $2.0\%$, and $1.7\%$ for $\chi_{cJ} \to\Sigma^0\bar{\Sigma}^0\eta$ with $J=0$, $1$, and $2$, respectively.

The systematic uncertainty associated with the fit includes uncertainties in the fit range, signal shape, background shape, peaking backgrounds, and the $\eta$ sideband scale factor. The uncertainty of the fit range is evaluated using the Barlow test~\cite{Barlow}. By adjusting the fit range ten times, for example, from (3290, 3590) $\mathrm{MeV}/c^2$ to (3310, 3610) $\mathrm{MeV}/c^2$ in 2 $\mathrm{MeV}/c^2$ steps. The changes in the final results are no more than  $2\sigma$, and therefore the corresponding systematic uncertainty is negligible. The uncertainty related to the signal shape, and the peaking background shape, is assessed by convolving the nominal signal shape with an additional Gaussian function, which represents the resolution difference between data and MC simulation. The discrepancy between the results obtained with the nominal and modified shapes is taken as the corresponding systematic uncertainty. Similarly, the uncertainty associated with the non-peaking background shape is estimated by replacing the second-order polynomial with a third-order polynomial, and the differences in the fitted signal yields are taken as the systematic uncertainties. Both the systematic uncertainties due to signal and background shapes are less than $0.1\%$ and are thus ignored. The contribution of potential peaking backgrounds is estimated by generating MC samples of $\psi(3686) \to \gamma \chi_{cJ}$ with  $\chi_{cJ} \to \eta \Sigma(1385) \bar{\Lambda}$ and $\Sigma(1385) \to \pi^0 \Lambda$. Assuming conservatively that the branching fraction of $\chi_{cJ} \to \eta \Sigma(1385) \bar{\Lambda} + c.c.$ is the same as that of $\chi_{cJ} \to \eta \Sigma^0 \bar{\Sigma}^0$, they are determined by comparing the results with and without considering the peaking background contribution, which are  $2.3\%$, $2.1\%$, and $2.1\%$ for $J=0$, $1$, and $2$, respectively. The systematic uncertainty arising from the $f_\eta$ scale factor is obtained by varying the $f_\eta$ value by one standard deviation, which is determined during the calculation of this factor, and comparing the differences in the final fit results. They are assigned to be $0.8\%$, $1.3\%$, and $0.3\%$ for $J=0$, $1$, and $2$, respectively.

Due to the presence of mis-combinations in the event reconstruction, the $M_{\Sigma^0\bar{\Sigma}^0\eta}$ distributions of the signal MC samples need to be fitted to determine the efficiencies. The resulting uncertainties in the fits are assigned as the corresponding systematic uncertainties, which are $0.5\%$, $0.4\%$, and $0.5\%$ for $J=0$, $1$, and $2$, respectively.

The nominal physics model of $\chi_{J} \to \Sigma^0\bar{\Sigma}^0\eta$ used in the MC simulation is the PHSP model. Its uncertainty is estimated by re-weighting the MC events with the distribution of $M(\Sigma \eta)$ obtained from the data after subtracting backgrounds using the $\eta$ sideband method and correcting for efficiency. The difference in efficiencies between the corrected and uncorrected MC samples is taken as the corresponding uncertainty. They are $0.9\%$, $2.8\%$, and $0.8\%$ for $J=0$, $1$, and $2$, respectively.

The uncertainties in the quoted branching fractions of $\psi(3686)\to\gamma\chi_{c0}$, $\psi(3686)\to\gamma\chi_{c1}$, $\psi(3686)\to\gamma\chi_{c2}$, $\Sigma^0\to\gamma\Lambda$, $\Lambda\to p \pi$, and $\eta\to\gamma\gamma$ are 2.3\%, 2.8\%, 2.5\%, $<0.1$\%, 0.8\%, and 1.0\%, respectively~\cite{pdg}.

The uncertainty related to the total number of $\psi(3686)$ events is 0.5\%~\cite{totalnumber_uncer}.

All relative systematic uncertainties are detailed in Table~\ref{TotalUncertainties}. The individual uncertainties are considered to be independent and are combined in quadrature, with the sum representing the total systematic uncertainty. 

\begin{table}[tbhp]
\begin{center}
\caption{The relative systematic uncertainties in the branching fraction measurements of $\chi_{cJ}\to\Sigma^0\bar{\Sigma}^0\eta$ (in $\%$). N means negligible. Here, “Brs quoted” indicates the uncertainties of the branching fractions quoted from PDG except the uncertainty of $\mathcal{B}(\psi(3686) \to \gamma \chi_{cJ})$. And “Total of prod” indicates the total systematic uncertainty of the product of the branching fractions $\mathcal{B}(\psi(3686)\to\gamma\chi_{\rm c J})\cdot\mathcal{B}(\chi_{\rm cJ} \to \Sigma^0\bar{\Sigma}^0\eta)$. }
\label{TotalUncertainties}
\begin{tabular}{l  c  c  c}
\\\hline \hline Source & $\chi_{c 0}$ & $\chi_{c 1}$ & $\chi_{c 2}$ \\
\hline  Photon reconstruction&  $2.5$    &   $2.5$   &  $2.5$   \\
Signal region &N &N &N\\
Veto &N &N &N\\
Kinematic fit &  1.8  &  2.0  &  1.7  \\
$\Lambda$ reconstruction & 1.8  & 1.8  & 1.8  \\
 $\bar{\Lambda}$ \text{reconstruction} & 1.5  & 1.5  & 1.5  \\
Fit range  & {N} & {N}  & {N}  \\
Signal shape & N  & N  & N  \\
Background shape & N  & N  & N  \\
 Peaking background &  2.3  &  2.1  &  2.1 \\
 {{Physics model}} & { 0.9 }  & { 2.4 }  & { 0.8 }\\

 {{Non-$\eta$ scale factor}} & { 0.8 }  & {1.3 }  & { 0.3}\\
 {{Mis-combination}} & { 0.5 }  & {0.4 }  & {0.5}\\
 Brs quoted &  1.3  &  1.3  &  1.3 \\
 $N_{\psi(3686)}$ & {{0.5}}  & {{0.5}}  & {{0.5}}  \\
\hline  
Total of prod  & {5.3}  & {{5.9}}  & {{5.1}}  \\
\hline \hline
$\mathcal{B}(\psi(3686) \to \gamma \chi_{cJ})$ & 2.3 & 2.8 & 2.5 \\
\hline
Total of $\mathcal{B}(\chi_{\rm cJ}\to\Sigma^0\bar{\Sigma}^0\eta)$  & {5.7}  & {{6.4}}  & {{5.5}}  \\
\hline \hline
\end{tabular}
\end{center}
\end{table}

	
\section{Summary and discussion}\label{sec:summary}
By analyzing $(2712.4\pm14.3)\times10^6 \ \psi(3686)$ events, the decays of $\chi_{cJ} \to \Sigma^0 \bar{\Sigma}^0 \eta$ for $J=$ 0, 1, and 2 are observed for the first time, with significances of 7.2$\sigma$, 5.7$\sigma$, and 6.7$\sigma$, respectively. Their branching fractions are determined to be $\mathcal B(\chi_{c0}\to\Sigma^0\bar{\Sigma^0}\eta) = (9.7 \pm 1.7 \pm 0.6)\times 10^{-5}$,  $\mathcal B(\chi_{c1}\to\Sigma^0\bar{\Sigma^0}\eta) = (4.9 \pm 1.1 \pm 0.3)\times 10^{-5}$ and $\mathcal B(\chi_{c2}\to\Sigma^0\bar{\Sigma^0}\eta) = (6.1 \pm 1.2 \pm 0.4)\times 10^{-5}$, where the first uncertainties are statistical  and the second systematic. No significant enhancement is detected near the $\Sigma^0\bar{\Sigma}^0$ production threshold. Additionally, no evidence of an excited $\Sigma^0$ state is observed in either the $M_{\Sigma^0\eta}$ or $M_{\bar{\Sigma}^0\eta}$ invariant mass distributions.

\begin{table}[H]
\begin{center}
\caption{The ratios of branching fractions $\mathcal B(other\ channel)/\mathcal B(\chi_{cJ}\to\Sigma^0\bar{\Sigma}^0\eta)$. The uncertainties include both systematic and statistical effects. The branching fractions of $\chi_{cJ}\to\Sigma^0\bar{\Sigma}^0\eta$ are from this analysis, the others are cited from Refs.~\cite{BESIII:2017szd,BESIII:2020raf,BESIII:2022rsg,BESIII:2024sfs}.}
\label{list10}
\begin{tabular}{l  c  c  c}
\hline \hline  Other channel & $\chi_{c 0}$ & $\chi_{c 1}$ & $\chi_{c 2}$ \\
\hline $\chi_{cJ} \to \Sigma^{+} \bar{\Sigma}^{-} $ &$0.50 \pm 0.10 $ &$0.07 \pm 0.02$ &$0.05 \pm 0.02$\\
 $\chi_{cJ} \to \Sigma^{0} \bar{\Sigma}^{0} $ &$0.47 \pm 0.09$ &$0.08 \pm 0.02$ &$0.06 \pm 0.01$\\
 $\chi_{cJ} \to \Sigma^{+} \bar{\Sigma}^{-} \eta$&  $1.26 \pm 0.33$    &   $1.39 \pm 0.37$   &  $0.88 \pm 0.27$   \\
 $\chi_{cJ} \to \Lambda \bar{\Lambda} \eta $ &$2.31 \pm 0.55$ &$1.20 \pm 0.42$ &$0.19 \pm 0.07$\\
\hline \hline
\end{tabular}
\end{center}
\end{table}

\begin{table}[H]
\begin{center}
\caption{The various ratios between $ \chi_{cJ}\to\Sigma^0\bar{\Sigma}^0\eta$ and $\chi_{c0}\to\Sigma^0\bar{\Sigma}^0\eta$ in this analysis. $\Omega$ are the three-body phase space factors. $\mathcal M$ are the accordingly determined amplitudes. The uncertainties include both statistical and systematic effects.}
\label{list11}
\begin{tabular}{l  c  c}
\hline \hline Ratio & $\chi_{c 1}$ & $\chi_{c 2}$  \\
\hline  $\mathcal B_{\chi_{c1,2}}/\mathcal B_{\chi_{c0}}$&  $0.50 \pm 0.14$    &   $0.63 \pm 0.17$      \\
$\Omega_{\chi_{c1,2}}/\Omega_{\chi_{c0}}$ &$1.38 $ &$1.58$ \\
$|\mathcal M_{\chi_{c1,2}}/\mathcal M_{\chi_{c0}}|^2$ &$0.36 \pm 0.10$ &$0.40 \pm 0.11$ \\
\hline \hline
\end{tabular}
\end{center}
\end{table}

We determine the ratios of the branching fractions for the decays of $\chi_{cJ}\to \Sigma^+ \bar{\Sigma}^-$, $\Sigma^0 \bar{\Sigma}^0$, $\Sigma^+ \bar{\Sigma}^- \eta$ and $\Lambda \bar{\Lambda} \eta$ that relative to $\chi_{cJ} \to \Sigma^0 \bar{\Sigma}^0 \eta$. The obtained results are summarized in Table~\ref{list10}. For the ratios $\mathcal B(\chi_{cJ}\to \Sigma\bar \Sigma)/\mathcal B( \chi_{cJ} \to \Sigma^0 \bar{\Sigma}^0 \eta)$, enhancements of the branching fractions of $\chi_{cJ} \to \Sigma^0 \bar{\Sigma}^0 \eta$  are observed. The ratios $\mathcal B( \chi_{cJ} \to \Sigma^+ \bar{\Sigma}^- \eta)/\mathcal B(\chi_{cJ} \to \Sigma^0 \bar{\Sigma}^0 \eta)$ are consistent with unity within one standard deviation, as expected based on isospin symmetry, which indicates that both decays primarily occur via strong interactions. However, no clear pattern is observed in  the ratios $\mathcal B( \chi_{cJ} \to \Lambda \bar{\Lambda} \eta)/\mathcal B(\chi_{cJ} \to \Sigma^0 \bar{\Sigma}^0 \eta)$, which may be due to large uncertainties. Additionally, Table~\ref{list11} presents the ratios of decay amplitudes for $\chi_{cJ} \to \Sigma^0 \bar{\Sigma}^0 \eta$. The $\mathcal B$ are the branching fractions. The $\Omega$ are the three-body phase space factors. The $\mathcal M$ are the accordingly determined amplitudes. The $|\mathcal M_{\chi_{c1,2}}/\mathcal M_{\chi_{c0}}|^2$ is calculated by ${\frac{\mathcal B_{\chi_{c1,2}}}{\mathcal B_{\chi_{c0}}}\cdot\frac{\Omega_{\chi_{c0}}}{\Omega_{\chi_{c1,2}}}}$. The results indicate that the amplitude for $\chi_{c0}$ differs from those of $\chi_{c1}$ and $\chi_{c2}$, implying distinct decay mechanisms or possible mixing between the $\chi_{c0}$ and an excited scalar glueball, the latter having a mass predicted by unquenched lattice QCD~\cite{Gregory:2012hu}. Further experimental measurements are required to elucidate the underlying dynamics.


{\section{Acknowledgement}}
The BESIII Collaboration thanks the staff of BEPCII (https://cstr.cn/31109.02.BEPC) and the IHEP computing center for their strong support. This work is supported in part by National Key R\&D 
Program of China under Contracts Nos. 2025YFA1613900, 2023YFA1606000, 2023YFA1606704; National Natural Science Foundation of China (NSFC) under Contracts Nos. 12375083, 11635010, 11935015, 11935016, 11935018, 12025502, 12035009, 12035013, 12061131003, 12192260, 12192261, 12192262, 12192263, 12192264, 12192265, 12221005, 12225509, 12235017, 12342502, 12361141819; the Chinese Academy of Sciences (CAS) Large-Scale Scientific Facility Program; the Strategic Priority Research Program of Chinese Academy of Sciences under Contract No. XDA0480600; CAS under Contract No. YSBR-101; 100 Talents Program of CAS; The Institute of Nuclear and Particle Physics (INPAC) and Shanghai Key Laboratory for Particle Physics and Cosmology; ERC under Contract No. 758462; German Research Foundation DFG under Contract No. FOR5327; Istituto Nazionale di Fisica Nucleare, Italy; Knut and Alice Wallenberg Foundation under Contracts Nos. 2021.0174, 2021.0299, 2023.0315; Ministry of Development of Turkey under Contract No. DPT2006K-120470; National Research Foundation of Korea under Contract No. NRF-2022R1A2C1092335; National Science and Technology fund of Mongolia; Polish National Science Centre under Contract No. 2024/53/B/ST2/00975; STFC (United Kingdom); Swedish Research Council under Contract No. 2019.04595; U. S. Department of Energy under Contract No. DE-FG02-05ER41374. 

	

\end{document}